\documentclass{iaus}
\usepackage{graphicx}

\title[Magnetic fields in spiral galaxies] 
{Interplay of CR-driven galactic wind, magnetic field, and galactic dynamo in spiral galaxies}

\author[M. Krause]   
{Marita Krause}

\affiliation{$^1$Max-Planck-Institut f\"ur Radioastronomie, Auf dem H\"ugel 69, 53121 Bonn, Germany \break email: mkrause@mpifr-bonn.mpg.de\\[\affilskip]
}

\pubyear{2009}
\volume{259}  
\pagerange{100--100}
\date{Nov. 20, 2008}
\jname{Cosmic Magnetic Fields: From Planets, to Stars and Galaxies}
\editors{K.G. Strassmeier, A.G. Kosovichev \& J.E. Beckman, eds.}
\begin{document}

\maketitle

\begin{abstract}
From our radio observations of the magnetic field strength and
large-scale pattern of spiral galaxies of different Hubble types
and star formation rates (SFR) we conclude that -- though a high
SFR in the disk increases the total magnetic field strength in
the disk and the halo -- the SFR does not change the global field
configuration nor influence the global scale heights of the radio
emission. The similar scale heights indicate that the total magnetic
field regulates the galactic wind velocities. The galactic wind
itself may be essential for an effective dynamo action.

\end{abstract}

\firstsection 
\section{Magnetic field strength and star formation}

Observations of a sample of three late-type galaxies with low surface-brightness and the radio-weak edge-on galaxy NGC~5907 (all with a low SFR) revealed that they all have an unusually high thermal fraction and weak total and regular magnetic fields (Chy{\.z}y et al. 2007, Dumke et al. 2000). However, these objects still follow the total radio-FIR correlation, extending it to the lowest values measured
so far. Hence, these galaxies have a lower fraction of synchrotron emission than galaxies with higher SFR. It is known that the thermal intensity is proportional to the SFR. Our findings fit to the equipartition model for the radio-FIR correlation (Niklas \& Beck 1997), according to which the nonthermal emission increases $\propto SFR^{1.3 \pm 0.2}$ and the \emph{total} magnetic field strength $ \rm B_t$ increases $\propto SFR^{0.34 \pm 0.14}$.\\
No similar simple relation exists for the \emph{regular} magnetic field strength. We integrated the polarization properties in 41 nearby spiral galaxies and found that (independently of inclination effects) the degree of polarization is lower ($ < 4\%$) for more luminous galaxies, in particular those for $ L_{4.8} > 2 \times 10^{21}~\rm{W Hz^{-1}}$ (Stil et al. 2008). The radio-brightest galaxies are those with the highest SFR. Though a dynamo action needs star formation and supernova remnants as the driving force for velocities in vertical direction, we conclude from our observations that stronger star formation seems to reduce the magnetic field regularity.
On kpc-scales, Chy{\.z}y (2008) analyzed the correlation between magnetic field regularity and SFR locally within one galaxy, NGC~4254. While he found that the total and random field strength increase locally with SFR, the regular field strength is locally uncorrelated with SFR.\\

\section{Vertical scale heights and CR-driven galactic wind}

We determined the exponential scale heights of the total power emission at $\lambda6$~cm for four edge-on galaxies (NGC~253, NGC~891, NGC~3628, NGC~4565) for which we have combined interferometer and single-dish data (VLA and the 100-m Effelsberg). In spite of their different intensities and extents of the radio emission, the vertical {\em scale heights} of the thin disk and the thick disk/halo are similar in this sample (300~pc and 1.8~kpc) (Dumke \& Krause 1998, Heesen et al. 2009). We stress that our sample includes the brightest halo observed so far, NGC~253, with a very high SFR, as well as one of the weakest halos, NGC~4565, with a small SFR.\

For NGC~253 Heesen et al. (this volume) argued that the synchrotron lifetime (which is $\propto \rm B_t ^{-2}$ ) mainly determines the vertical scale height of the synchrotron emission and estimated the cosmic ray bulk velocity to $300 \pm 30$~km/s. As this is similar to the escape velocity, it shows the presence of a galactic wind in this galaxy. The fact that we observe similar averaged scaleheights at $\lambda6$~cm for the
four galaxies mentioned above imply that the galactic wind velocity is proportional to $\rm B_t ^2$, and hence proportional to $\rm {SFR} ^{0.7 \pm 0.3}$.\

\section{Magnetic field structure, dynamo action, and galactic wind}

In a larger sample of 11 edge-on galaxies we found in all of them (except the inner part of NGC~4631, see Krause 2009) mainly a disk-parallel magnetic field along the galactic midplane together with an X-shaped poloidal field in the halo. Our sample includes spiral galaxies of different Hubble types and SFR, ranging from $0.5~\rm{M}_\odot \rm{yr}^{-1} \le \rm{SFR} \le 27~\rm{M}_\odot \rm{yr}^{-1}$. The disk-parallel magnetic field is the expected edge-on projection of the spiral magnetic field within the disk as observed in face-on galaxies. It is generally thought to be generated by a mean-field $\alpha \Omega$-dynamo for which the most easily excited field pattern is the axismmetric spiral (ASS) field (e.g. Beck et al. 1996). The poloidal part of the ASS dynamo field alone, however, cannot explain the observed X-shaped structures in edge-on galaxies as the field strength there seems to be comparable to that of the large-scale disk field. Model calculations of the mean-field $\alpha\Omega$-dynamo for a disk
surrounded by a spherical halo including a {\em galactic wind} (Brandenburg et al. 1993) simulated similar field configurations as the observed ones. New MHD simulations are in progress (see e.g. Gressel et al. this volume, Hanasz et al. this volume) which include a galactic wind implicitely. A galactic wind can also solve the helicity problem of dynamo action (e.g. Sur et al. 2007). Hence, a galactic wind may be essential for an effective dynamo action, and to explain the observed similar vertical scale heights and X-shaped magnetic field structure in edge-on galaxies.

\end{document}